\documentclass[twocolumn,english,pra]{revtex4}
\usepackage[T1]{fontenc}
\usepackage[latin9]{inputenc}
\usepackage{graphicx}
\usepackage{esint}

\makeatletter
\@ifundefined{textcolor}{}
{%
 \definecolor{BLACK}{gray}{0}
 \definecolor{WHITE}{gray}{1}
 \definecolor{RED}{rgb}{1,0,0}
 \definecolor{GREEN}{rgb}{0,1,0}
 \definecolor{BLUE}{rgb}{0,0,1}
 \definecolor{CYAN}{cmyk}{1,0,0,0}
 \definecolor{MAGENTA}{cmyk}{0,1,0,0}
 \definecolor{YELLOW}{cmyk}{0,0,1,0}
 }

\makeatother

\usepackage{babel}

\begin{document}

\title{Ferro-electric phase transition in a polar liquid and the nature
of $\lambda-$transition in supercooled water. }

\author{P.O. Fedichev$^{1,2}$ and L.I. Menshikov$^{1}$}

\affiliation{$^{1}$Russian Research Center {}``Kurchatov Institute'', Kurchatov
Square 1, Moscow, 123182, Russia}

\affiliation{$^{2}$Quantum Pharmaceuticals Ltd, Ul. Kosmonavta Volkova 6A-606,
Moscow, 125171, Russia}
\begin{abstract}
We develop a series of approximations to calculate free energy of
a polar liquid. We show that long range nature of dipole interactions
between the molecules leads to para-electric state instability at
low temperatures and to a second-order phase transition. We establish
the transition temperature, $T_{C}$, both within mean field and ring
diagrams approximation and show that the ferro-electric transition
may play an important role explaining a number of peculiar properties
of supercooled water, such as weak singularity of dielectric constant
as well as to a large extent anomalous density behavior. Finally we
discuss the role of fluctuations, shorter range forces and establish
connections with phenomenological models of polar liquids.
\end{abstract}
\maketitle
Water is a major and all-important example of a strongly interacting
polar liquid. Dielectric properties of water surrounding nano-scale
objects pose a fundamentally important problem in physics, chemistry,
structural biology and in silica drug design. The issue of temperature
dependence of dielectric constant, the role of fluctuations and a
possibility of a ferro-electric phase transition in a polar liquid
is fairly old \citep{debye1929polar}. It attracted a new attention
when a new phase transition (so called $\lambda-$transition) was
observed in supercooled water at critical temperatures between $T_{c}=228^{0}K$
\citep{angell1973anomalous,hodge1978relative,speedy1976isothermal}
and $T_{c}=231.4^{0}K$ \citep{ter1981thermodynamic}. Isothermal
compressibility, density, diffusion coefficient, viscosity and static
dielectric constant, $\epsilon$, and other quantities show to a different
degree singular behavior $T_{c}$ is approached, which is signature
of a phase transition. Phase behavior of such metastable water has
been studied in a great detail by different authors a number of different
but apparently interrelated theories where suggested. One popular
view is given by the so called {}``stability limit conjecture'':
thermodynamic anomalies arise from a single limit of mechanical stability
(spinodal line), originating at the liquid-gas critical point \citep{speedy1976isothermal,speedy1982limiting,speedy1982stability}.
Molecular dynamics shows that supercooling anomalies may be caused
by a newly identified critical point, above which the two metastable
amorphous phases of ice become indistinguishable \citep{poole1992phase}.
In \citep{stillinger1980water} the phase transition is explained
as a formation of a rigid network of hydrogen bonds. One the other
hand, the singularity of $\epsilon$ is a feature of a ferro-electric
transition \citep{angell1983sw,stillinger1977tai}. A ferro-electric
hypothesis was also proposed and supported by molecular-dynamics simulations
(MD). For example, a ferro-electric liquid phase was observed in a
model of the so called {}``soft spheres'' with static dipole moments
\citep{wei1992flc,weis1992oos,groh1994fps,groh1994lro}. In fact,
the existence of a ferro-electric phase appears to be model independent:
domains where formed both in MD calculations with hard spheres with
point dipoles \citep{weis1992oos,matyushov2007model} and with soft
spheres with extended dipoles \citep{ballenegger2003sad}.

In this Letter we develop two related approaches to calculate free
energy of a polar liquid. We show that long range nature of dipole
interactions between the molecules leads to para-electric state instability
at sufficiently low temperatures and to a second-order phase transition.
We establish the transition temperature, $T_{c}$, both within mean
field and ring diagrams approximation and demonstrate that the ferro-electric
transition is a sound physical explanation behind the experimentally
observed $\lambda-$transition in supercooled water. Finally we discuss
dielectric properties, the role of fluctuations and establish connections
with earlier phenomenological models \citep{fedichev2006long,gong2009langevin}
of polar liquids.

Let us start with potential energy of a model system of dipoles (molecules)
located at positions $\mathbf{r_{i}}$ and numbered $i=1,2,...N$:
\begin{equation}
U=\frac{1}{2}\sum_{ij}u_{ij}.\label{eq:EnergyFull}\end{equation}
The pair interactions $u_{ij}$ between the molecules $i$ and $j$
forming the liquid can be modeled as \begin{equation}
u_{ij}=\frac{f(r_{ij})d_{0}^{2}}{\epsilon_{\infty}}\sum_{\alpha,\beta}(\mathbf{S}_{i})_{\alpha}(\mathbf{S}_{j})_{\beta}[\delta_{\alpha\beta}-3(\hat{\mathbf{r}_{ij}})_{\alpha}(\hat{\mathbf{r}_{ij}})_{\beta}],\label{eq:UDipoleDipole}\end{equation}
where $r_{ij}=|{\mathbf{r}}_{ij}|$ stand for the vector separations
between the molecules: $\mathbf{r}_{ij}=\mathbf{r}_{i}-\mathbf{r}_{j}$,
$\mathbf{d}_{i}=d_{0}\mathbf{S}_{i}$ is the static dipole moment
of a molecule $i$, $d_{0}$ is its absolute value and $\mathbf{S}_{i}$
is the unit vector in the direction of the dipole moment (summation
is assumed over repeated Cartesian indexes $\alpha$ and $\beta$).
The function $f(r_{ij})$ represents the spatial dependence of the
interaction between the molecules (see below). At last, $\epsilon_{\infty}$
is a dielectric constant of the liquid at frequencies exceeding rotational
frequencies of the molecules. In fact $\epsilon_{\infty}$ stands
for the electronic degrees of freedom polarization of the liquid molecules
and thus is not directly included in Eq.(\ref{eq:EnergyFull}).

At large distances $f(r_{ij})\approx1/r_{ij}^{3}$ as it should be
for a dipole-dipole interaction. At smaller distances the interaction
between the molecules is no longer described by its long-distance
approximation (\ref{eq:UDipoleDipole}). To account for the hydrogen
bonding and other short-distance effects (here and on for the concreteness
we should speak about water) without loosing a possibility to perform
analytical calculations we take the following, approximate, representation
for the intermolecular interaction:\[
f(r_{ij})=\left\{ {1/r_{ij}^{3},\; r_{ij}>r_{0}\atop 0,\; r_{ij}\leq r_{0}},\right.\]
where $r_{0}$ is model-dependent characteristic scale of interaction
between molecules (see e.g. \citep{frodl1992bai}). In spite of the
simplicity of the suggested approximation for the intermolecular interaction,
large scale properties of the liquid do not depend on exact value
of $r_{0}$ (see below).

The dipole-dipole interaction falls faster with $r_{ij}$ than the
Coulomb one. Nevertheless it should be also considered as a long-range
interaction. Indeed, let us consider the case of a uniformly polarized
medium $\langle\mathbf{S}_{i}\rangle=\mathbf{s}(\mathbf{r})={\rm {\rm const}}$.
Within the mean field approximation (MFA) the mean electric field
at the position of the molecule $i$ equals \citep{Stratton,jackson1999ce,martin2008cavity}
\begin{equation}
\mathbf{E}_{M}(\mathbf{r}_{i})=\sum_{j}\frac{f(r_{ij})}{\epsilon_{\infty}}(\mathbf{d}_{j}-3\hat{\mathbf{r}}_{ij}(\hat{\mathbf{r}}_{ij},\mathbf{d}_{j}))=\mathbf{E}_{A}+\mathbf{E}_{P},\label{eq:MoleculeField}\end{equation}
where $\mathbf{E}_{A}=4\pi\mathbf{P}/3/\epsilon_{\infty}$ is the
so called {}``action field'', $\mathbf{P}=P_{0}\mathbf{s}$ is the
macroscopic polarization vector \citep{jackson1999ce}, and $P_{0}=n_{0}d_{0}$
($n_{0}$ is the particle density of the liquid). $\mathbf{E}_{P}$
is the polarization electric field produced by the polarization charges
with a surface density $\sigma=P_{n}$, which may be as far away as
on the sample surface. 

The calculation of the free energy can be done with the help of thermodynamic
integration: let us {}``switch on'' the pair interaction between
the molecules, $U\rightarrow\lambda U$, adiabatically by changing
the scaling factor $\lambda$ from $0$ to $1$. Then, the correlation
energy, $F_{corr}=F-F_{0}$, where $F_{0}$ is the energy of the non-interaction
system, can be obtained by the formal integration \citep{LandauLifshitzStatPhys5}:
\begin{equation}
F_{corr}=\int_{0}^{1}d\lambda\langle U\rangle.\label{eq:LambdaWay}\end{equation}
The averaging $\langle...\rangle$ here and everywhere below is assumed
to be thermodynamic average over the Gibbs distribution:\[
\langle U\rangle=Z^{-1}\int d\gamma U\exp(-\lambda U/T),\]
where $d\gamma=\prod_{i}d^{3}x_{i}d\Omega_{\mathbf{S}_{i}}/4\pi V$,
$Z=\int d\gamma\exp(-\lambda U/T),$ and $V$ is the volume of the
system ($n_{0}V=N$). 

\begin{figure}
\includegraphics[width=0.8\columnwidth]{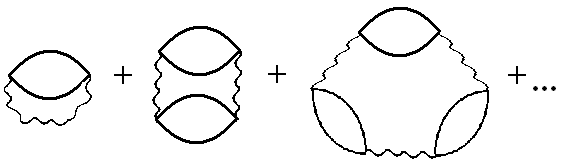}

\caption{Ring diagrams approximation to free energy of an interacting system
(\ref{eq:SumOfRingDiagrams}): Wavy lines and polarization loops represent
electrostatic dipole-dipole interaction (\ref{eq:UDipoleDipole})
and averaging over molecular degrees of freedom, respectively \citep{ll:volIX}.
\label{fig:Matsubara}}

\end{figure}

As it is well known in plasma physics and will be further elucidated
here for our dipole liquid example, MFA to the free energy can be
obtained with a help of the so called ring diagrams approximation
(RA) \citep{ll:volIX}, as shown on Figure \ref{fig:Matsubara}:\begin{equation}
\frac{\partial F}{\partial\lambda}=\sum_{n=0}^{\infty}\frac{\partial F^{(n)}}{\partial\lambda},\label{eq:SumOfRingDiagrams}\end{equation}
where \begin{equation}
\frac{\partial F^{(n)}}{\partial\lambda}=\frac{1}{2}\left(\frac{-\lambda}{T}\right)^{n}\sum_{i_{1}...i_{n+1}}\int d\gamma_{i_{1}}...d\gamma_{i_{n+1}}u_{i_{1}i_{2}}...u_{i_{n+1}i_{1}}.\label{eq:RingDiagramNthOrderRSpace}\end{equation}
Applying Fourier transform and integrating over directions $d\Omega_{\mathbf{S}}$
of the molecular dipole moments we find that \[
f(r)(\delta_{\alpha\beta}-3\hat{r}_{\alpha}\hat{r}_{\beta})=\int\frac{d^{3}k}{(2\pi)^{3}}B(k)L_{\alpha\beta}(\hat{k})\exp(i\mathbf{kr}),\]
where $L_{\alpha\beta}(\hat{k})=\delta_{\alpha\beta}-3\hat{k}_{\alpha}\hat{k}_{\beta},$
and $B(k)=4\pi(z\cos z-\sin z)/z^{3}$, we find, that: \[
\frac{\partial F^{(n)}}{\partial\lambda}=\frac{V}{2}\left(\frac{-\lambda}{T}\right)^{n}\left(\frac{d_{0}^{2}N}{3\epsilon_{\infty}V}\right)^{n+1}\int\frac{dk^{3}}{(2\pi)^{3}}\left(B(k)\right)^{n+1}Q_{n},\]
where $Q_{n}=2\left[1-2^{n}(-1)^{n}\right]$. The summation of all
the ring diagrams and integration over $\lambda$ finally lead to:
\[
F_{corr}=T\int\frac{Vd^{3}k}{(2\pi)^{3}}\left[\ln\left(1+\frac{n_{0}d_{0}^{2}B(k)}{3\epsilon_{\infty}T}\right)+\right.\]
\begin{equation}
\left.+\frac{1}{2}\ln\left(1-\frac{2n_{0}d_{0}^{2}B(k)}{3\epsilon_{\infty}T}\right)\right].\label{eq:FcorrDDcomplete}\end{equation}
At $kr_{D}\ll1$, $B(k)\approx-4\pi/3+2\pi(kr_{0})^{2}/15$, and the
first term in Eq. (\ref{eq:FcorrDDcomplete}) takes the form: \begin{equation}
F_{corr}^{(1)}=T\int\frac{Vd^{3}k}{(2\pi)^{3}}\ln\left(1-\frac{T_{c}}{T}+\gamma(kr_{0})^{2}\right),\label{eq:FcorrDDFinal}\end{equation}
where $\gamma=1/10$ and \begin{equation}
T_{c}=\frac{4\pi n_{0}d_{0}^{2}}{9\epsilon_{\infty}}.\label{eq:Tc}\end{equation}
Of course, the parameter $\gamma$ is model-dependent, and depends
on the details of short-range interactions between the molecules.
Although its value for a specific liquid is not really known, we expect
$\gamma\sim1$ (see the connection to a polar liquid phenomenology
below). 

Free energy (\ref{eq:FcorrDDFinal}) of the liquid has a meaning only
at sufficiently high temperatures: $\tau=(T-T_{c})/T_{c}>0$. At $T<T_{c}$
the free energy (\ref{eq:FcorrDDFinal}) becomes singular, the para-electric
phase ($\mathbf{s}=0$) becomes unstable and a ferro-electric phase
transition takes place at $T=T_{c}$.

Below the phase transition, at $T<T_{c}$, a simpler approach can
be taken. Indeed in thermodynamic equilibrium the density of polarization
charges vanishes: $\rho_{P}=-\mathbf{\nabla P}=0$ due to the formation
of a domains (see \citep{ll:volVI}) and, therefore, $\mathbf{E}_{P}=0$
as well. Let us neglect fluctuations first and consider a polarized
state of a liquid with $\mathbf{E}_{P}=0$. Then, the energy of the
liquid is\begin{equation}
U=\frac{1}{2}\sum_{i}\mathbf{d}_{i}\mathbf{E}_{M}(\mathbf{r}_{i})=-\frac{2\pi n_{0}d_{0}^{2}Ns^{2}}{3\epsilon_{\infty}}.\label{eq:UsHomogeneous}\end{equation}
In the vicinity of the phase transition, $|\tau|\ll1$, the polarization
of the liquid is small: $s\ll1$. One-particle distribution function
for the molecular dipoles orientation has the form: $f(\theta)=(1+3s\cos\theta)/4\pi$,
where $\theta$ is the angle between vectors $\mathbf{S}$ and $\mathbf{s}$.
Then $\langle s_{z}\rangle=s$ and entropy of the liquid is:\[
S=-N\int d\Omega_{\mathbf{S}}f\ln f\approx-\frac{3}{2}Ns^{2}-\frac{1}{15}Ns^{4}.\]
Combining the above expressions we obtain the following representation
of the non-equilibrium free energy\begin{equation}
F(\mathbf{s})=U-TS=as^{2}+bs^{4},\label{eq:FLandauGinzburg}\end{equation}
where $a=3N(T-T_{c})/2$ and $b=NT_{c}/15.$ The free energy (\ref{eq:FLandauGinzburg})
has the Landau form \citep{LandauLifshitzStatPhys5}, suggesting the
phase transition is of the second order at $T=T_{c}$. 

In a weak electric field $\mathbf{E}$ the free energy is $F\approx as^{2}-Nd_{0}\mathbf{sE}-N\mathbf{d}_{e}\mathbf{E}$,
where $\mathbf{d}_{e}=(\epsilon_{\infty}-1)\mathbf{E}/4\pi n_{0}$
is the induced electronic dipole moment of the molecules. At $T>T_{c}$
the minimization of $F$ gives $\mathbf{s}=\mathbf{E}d_{0}/3T_{c}\tau.$
Therefore the static dielectric constant is given by: \begin{equation}
\epsilon=\epsilon_{\infty}\left(1+\frac{3T_{c}}{T-T_{c}}\right)\label{eq:epsilonT}\end{equation}
and diverges at $T=T_{c}$. There are a few measurements of $\epsilon_{\infty}$
for water: $\epsilon_{\infty}\approx4.9$ \citep{stogrin1971} and
$\epsilon_{\infty}=5.5$ \citep{Zatsepina}. Plugging the experimental
values into Eq. (\ref{eq:Tc}) we obtain: $T_{c}=236^{0}K$ and $T_{c}=210^{0}K$,
respectively. The second order phase transition was indeed observed
in the supercooled water at $228^{0}K$ (\textquotedblleft{}the $\lambda$-transition\textquotedblright{}\citep{angell1973anomalous,speedy1976isothermal,hodge1978relative}),
which is in a good agreement with our findings. 

At $T\rightarrow T_{c}+0$ such values as isothermal compressibility,
density, diffusion coefficient of the molecule, viscosity and static
dielectric constant $\epsilon$ diverge. The singularity of $\epsilon$
led to ferro-electric phase transition hypotheses \citep{angell1983sw,stillinger1977tai}.
Eq. (\ref{eq:epsilonT}) shows that $\epsilon$ diverges as $\epsilon\propto|\tau|^{-1}.$
According to the experiments $\epsilon\propto|\tau|^{-0.13}$\citep{hodge1978relative}.
The reason of the discrepancy should be contributed to MFA failure
in the direct vicinity of the transition. At $T=0^{0}C$ depending
on the accepted value of $\epsilon_{\infty}$ we obtain $\epsilon=99$
and $\epsilon=61$, respectively, both close the observed value $\epsilon=88$
\citep{stogrin1971}. 

Finally let us connect our MFA model with polar liquid phenomenology
\citep{fedichev2006long}. Consider a more general case, $\mathbf{E}_{P}\neq\mathbf{0}$,
$\mathbf{s}(\mathbf{r})\neq{\rm const},$ $s\ll1$. Then, instead
of Eq. (\ref{eq:UsHomogeneous}) we obtain: \[
U=\frac{1}{2}\sum_{i\neq j}u_{ij}=-\frac{1}{2}\sum_{i}\mathbf{d}_{i}(\mathbf{E}_{A}(\mathbf{r}_{i})+\mathbf{E}_{P}(\mathbf{r}_{i})).\]
Taking into account Poisson equation, $\Delta\phi_{P}=-4\pi\rho_{P}/\epsilon_{\infty}$,
the definition for the density of polarization charges, $\rho_{P}=-P_{0}{\rm div}\mathbf{s}$,
and the electrostatic potential, $\mathbf{E}_{P}=-\mathbf{\nabla}\phi_{P}$,
we obtain:\[
U=\int dV\left(-\frac{2\pi P_{0}^{2}}{3\epsilon_{\infty}}s^{2}(\mathbf{r})+\frac{\epsilon_{\infty}\mathbf{E^{2}}_{P}}{8\pi}\right).\]
The entropy of the liquid can be expressed as \[
S=n_{0}\int dV\left(-\frac{3}{2}s^{2}(\mathbf{r})-\frac{1}{15}s^{4}(\mathbf{r})\right).\]
Transition from homogeneous to inhomogeneous polarization cases means
the deformation of hydrogen bonds. Therefore, the MFA free energy
of the liquid should allow additional Oseen like positive {}``$H$-bonding''
term (see e.g. \citep{LandauLifshitzStatPhys5,fedichev2006long})
\begin{equation}
F[\mathbf{s}]=\int dV\left[\frac{P_{0}^{2}}{2}\left(C\sum_{\alpha,\beta}\frac{\partial s_{\alpha}}{\partial x_{\beta}}\frac{\partial s_{\alpha}}{\partial x_{\beta}}+V(s^{2})\right)+\frac{\epsilon_{\infty}\mathbf{E^{2}}_{P}}{8\pi}\right],\label{eq:FreeEnergyPhenom}\end{equation}
where $C$ is the phenomenological parameter responsible for the h-bonds
network rigidity, $V(s^{2}\ll1)\approx As^{2}/2+Bs^{4}$, and $A=4\pi\tau/3\epsilon_{\infty}$,
$B=4\pi/135\epsilon_{\infty}$. 

Eq. (\ref{eq:FreeEnergyPhenom}) with a general potential $V(s^{2})$
can be considered as a phenomenological free energy of a polar liquid.
At $s\sim1$ the function takes into account the short-range part
of the intermolecular interaction potential and thus depends on details
of interactions of individual molecules. Asymptotically $V^{\prime}(s^{2}\rightarrow1)\rightarrow\infty$
\citep{fedichev2006long}, although the exact shape of $V(s^{2})$
can only be established either from experiments or detailed MD studies.

Next to the phase transition, $|\tau|\ll1$, fluctuations become essential.
The free energy of a liquid can be obtained in a usual way \citep{LandauLifshitzStatPhys5}:\begin{equation}
F=-T\ln\left[\int D\mathbf{s}(\mathbf{r})\exp\left(-\frac{F(\mathbf{s}(\mathbf{r}))}{T}\right)\right],\label{eq:FuncIntegral4F}\end{equation}
i.e. as a result of averaging over fluctuations. Neglecting the nonlinear
part of $V(s^{2})$ and Fourier transforming the functional (\ref{eq:FreeEnergyPhenom})
we find that\[
F[\mathbf{s}]=\frac{P_{0}^{2}}{2}\sum_{\mathbf{k}}\left[(Ck^{2}+A)|\mathbf{s_{k}}|^{2}+4\pi|\hat{\mathbf{k}}\hat{\mathbf{s}}_{\mathbf{k}}|^{2}\right].\]
The integration in (\ref{eq:FuncIntegral4F}) is then straightforward
and gives:\begin{equation}
F=VT\int\frac{d^{3}k}{(2\pi)^{3}}\ln\left(1-\frac{T_{c}}{T}+\frac{3\epsilon_{\infty}C}{4\pi}k^{2}\right)+{\rm const}.\label{eq:FfromFluctuations}\end{equation}
The latter expression should be compared with Eqs. (\ref{eq:FcorrDDcomplete})
and (\ref{eq:FcorrDDFinal}) obtained within the ring approximation.
Since Eqs. (\ref{eq:FcorrDDFinal}) and (\ref{eq:FfromFluctuations})
have the same form, we imply that averaging in RA represents averaging
over fluctuations on top of MFA thermal state. Moreover we establish
the relation between the model-dependent {}``microscopic'' cutoff
parameter $r_{0}$ used in RA and the phenomenological constant $C$:
$\gamma r_{0}^{2}=3\epsilon_{\infty}C/4\pi.$ 

The water density maximum at $T=4^{0}C$ \citep{despretz1837odo,zheleznyi1969dsw,rasmussen1973csw,angell1983sw}
is often related to supercooled anomalies as well \citep{angell1981wct}.
Let us show that this density behavior in our model originates from
the dipole field fluctuations in the vicinity of the ferro-electric
transition. Consider the liquid approaching the phase transition from
above: $\tau\ll1,\; T>T_{c}$. In the region where the fluctuations
are still small, $s\ll1$, the complete Free energy takes form: 

\begin{equation}
F=\int dVd_{0}^{2}n^{2}\left[\frac{C}{2}C\left(\nabla_{\alpha}s_{\beta}\right)^{2}+V(s)\right]+\int dV\frac{\beta}{2}\left(n^{\prime}\right)^{2}\label{eq: Free energy of s and n fluctuations}\end{equation}
where $n^{\prime}(\mathbf{r})=n(\mathbf{r})-n_{0}$ represent the
density fluctuations, $\beta=mC_{S}^{2}/n_{0}$, $m$ is the mass
of the molecule, $C_{S}$ is the speed of a sound. Since the most
important contribution to $F$ comes from the long wave length fluctuations
of $\mathbf{s}$, the vector field $\mathbf{s}$ is slow variable
and therefore the density can be found by the minimization of the
free energy at a given value of $\mathbf{s}$:\begin{equation}
n^{\prime}=-\frac{P_{0}^{2}}{n_{0}\beta}\left[C\left(\nabla_{\alpha}s_{\beta}\right)^{2}+As^{2}+2Bs^{4}\right].\label{eq:Eq for density}\end{equation}
Averaging over the polarization field fluctuations we find that $\langle n^{\prime}\rangle=n_{1}+n_{2}$,
$n_{1}$ practically does not depend on $T$, and the temperature
dependence is contained in $n_{2}$: \begin{equation}
\langle n^{\prime}\rangle={\rm const}+n_{0}D\sqrt{\tau},\label{eq: final result for n sub two}\end{equation}
where $D=\xi Q$, $Q=BT^{2}k_{max}\epsilon_{\infty}^{2}/mC_{S}^{2}P_{0}^{2}R_{D}^{5}\sim1,$and
$R_{D}=\sqrt{C\epsilon_{\infty}/4\pi}$ is the correlation radius
introduced in \citep{fedichev2006long}. $R_{D}$ is roughly a size
of a tightly correlated domain of molecular dipoles and plays the
role similar to Debye radius in plasma models. The factor $\xi=5\left(2\pi\right)^{-5}/\sqrt{3}\ll1$
so that $D$ appears to be small. In fact the smallness of $\xi$
is the defect of the mean field theory and in a full theory of the
phase transition taking into account the scale invariant fluctuations
of the order parameter $\xi\sim1$, as shown in \citep{ginzburg1960srp,ginzburg1976shi}.
Eq. (\ref{eq: final result for n sub two}) shows that the density
of the liquid drops near the phase transition point. At larger temperatures
the density decreases due to a normal thermal expansion of the liquid.
Therefore complete model predicts maximum density at a certain temperature. 

Fluctuations in a homogeneous liquid become important when $C_{FL}=-T\partial^{2}F/\partial T^{2}\agt\Delta C$
\citep{ginzburg1960srp,vaks1966msc}, where $C$ and $\Delta C$ are
heat capacity and the jump of the heat capacity at $T=T_{c}$. The
application of this condition to the model (\ref{eq:FfromFluctuations})
gives the following range of fluctuation region: $|\tau|\alt1/z^{2}$,
where $z=4\pi n_{0}R_{D}^{3}/3.$ For water we have $R_{D}\sim0.3nm$
and $z\sim4$. This means the phenomenological (MFA) free energy (\ref{eq:FreeEnergyPhenom})
can be applied at $T\agt-20^{0}C$. The mean field approach is applicable,
since the number of particles within $R_{D}$ is large: $z\approx4\agt1.$

The combination of RA and MFA approaches together with the established
connection with polar liquid phenomenology explains a number of important
features behind the phase transition observed in supercooled water
at $T\approx-45^{0}C$ \citep{angell1973anomalous,speedy1976isothermal,hodge1978relative,ter1981thermodynamic}.
The {}``gaseous parameter'' $z$ for water is not very large, which
means that MFA approach to the phase transition developed here can
not explain every experimental or MD feature in a realistic situation.
Although our approximation (\ref{eq:Tc}) for $T_{c}$ and (\ref{eq:epsilonT})
for the dielectric constant turns out to be fairly good and appear
to capture all the important physics, calculations of quantities such
as critical exponents near $T_{c}$ should require a more complete
theory including all the details of short-distance hydrogen-bonds
network formation \citep{stillinger1977tai,stillinger1980water,sokolov1995dsw}
as well as scale invariant character of order parameter fluctuations
in the vicinity of the phase transition \citep{LandauLifshitzStatPhys5}.
In any case we believe that long range dipole-dipole interaction of
molecules plays a key role in explaining supercooled water anomalies
and interactions in aqueous environments \citep{fedichev2006long,gong2009langevin}.

The authors are indebted to Quantum Pharmaceuticals for support. The
polar liquid phenomenology backed by our research is employed in Quantum
Pharmaceuticals models and drug discovery applications.

\bibliographystyle{apsrev}
\bibliography{/home/fedichev/Documents/QS/Qrefs}

\end{document}